\documentclass[usenatbib,longnamesfirst]{mn2e}
\RequirePackage{graphicx}
\begin{document}
\title[Observations of Bok globules]{Optical and submillimetre observations of 
Bok globules -- tracing the magnetic field from low to high density}
\author[Ward-Thompson {\em et al.}]
{D. Ward-Thompson$^1$\thanks{e-mail: D.Ward-Thompson@astro.cf.ac.uk},
A. K. Sen$^2$, J. M. Kirk$^1$, D. Nutter$^1$ \\
$^1$School of Physics and Astronomy, Cardiff University, 
Cardiff, CF24 3AA\\
$^2$Department of Physics, Assam University, Silchar, 788 011, India}

\maketitle

\begin{abstract}
We present optical and submillimetre polarimetry data of the Bok globule CB3. We also present optical polarimetry and submillimetre continuum data of the Bok globule CB246. We use each set of polarimetry data to infer the magnetic field orientation in each of the clouds. The optical data can only  trace the field orientation in the low density edge  regions of clouds, because if the extinction is too high then no optical emission is transmitted. The submillimetre data can only trace the field orientation in the high density central regions of the clouds, because current submillimetre polarimeters are only sensitive to high column densities. 

It has previously been found that near-infrared polarisation mapping  of background stars does not accurately trace the magnetic field in dense cloud regions, and hence that the grains responsible for near-infrared polarisation are under-represented in those regions. This may be due to a lack of aligned grains in dense regions. We test this by comparing the field orientations measured by our two independent methods of optical and submillimetre polarimetry. We find that the field orientation deduced from the optical data matches up well with the orientation estimated from the submillimetre data. We therefore claim that both methods are accurately tracing the same magnetic field in CB3. Hence, in this case, there must be significant numbers of aligned dust grains in the high density region, and they do indeed trace the magnetic field in the submillimetre. 

We find an offset of 40$\pm$14 degrees between the magnetic field orientation and the short axis of the globule. This is consistent with the mean value of 31$\pm$3 degrees found in our previous work on prestellar cores, even though CB3 is a protostellar core. CB246 is a prestellar core, and in this case the offset between the magnetic field orientation inferred from the optical polarisation data and the short axis of the core inferred from the submillimetre continuum data is 28$\pm$20 degrees.  Taken together, the six prestellar cores that we have now studied in this way show a mean offset between magnetic field orientation and core short axis of $\sim$30$\pm$3 degrees,  in apparent contradiction with some models of magnetically dominated star formation.

\end{abstract}


\section{Introduction}\label{intro}

Many models of star formation predict that magnetic fields play some
role in the star formation process 
\citep{1956MNRAS.116..503M,1976ApJ...206..753M,1976ApJ...207..141M,
1978PASJ...30..671N,1994ApJ...425..142C,1999ApJ...513..259O,
2004ApJ...609L..83L}. However, measuring magnetic
fields in space is notoriously difficult. Zeeman measurements allow one
to measure the line-of-sight field strength using the line splitting of 
different electronic magnetic moment states in the presence of a magnetic
field \citep{1897ApJ.....5..332Z,1969ApJ...156..861V,1999ApJ...520..706C,
2008ApJ...680..457T}. 
Faraday rotation of a background polarized source seen through
a region containing a magnetic field can also be used to measure the
line-of-sight field strength through that region \citep{1966MNRAS.133...67B,
1998MNRAS.299..189S}.

The orientation of the magnetic field in the plane of the sky can be measured
via the effects of aligned dust grains. The method of grain alignment has
been a matter of debate for many years \citep[e.g.][]{1951ApJ...114..206D,
1967ApJ...147..943J,1979ApJ...231..404P,1995ApJ...451..660L,
1995MNRAS.277.1235L,
1996ApJ...470..551D}. However, the various
methods are generally lumped together under the name of the Davis-Greenstein 
effect \citep{1951ApJ...114..206D}. 

All methods predict that the grains align with their long axes
perpendicular to the magnetic field. Hence in the optical and 
near-infrared, where the
polarization is caused by preferential extinction
by aligned elongated grains of a background
unpolarized source \citep{1949Sci...109..166H,1949Sci...109..165H}, 
the polarization
orientation is parallel to the field. In the far-infrared and submillimetre,
where the polarization is caused by preferential emission
from aligned elongated dust grains \citep{1977ApJ...215..529D,
1982MNRAS.200.1169C,1984ApJ...284L..51H}, the
polarization orientation is perpendicular to the field orientation 
\citep{1977ApJ...215..529D}.

The method of tracing the magnetic field using
optical polarization of background starlight typically only
works in regions of low extinction (A$_V$ $\sim$ 1--5). 
So this technique is mainly applicable towards the low-density edge 
regions of the clouds. 
This is because these data use preferential
extinction of background starlight by aligned grains
to produce the polarisation. At higher
extinctions the background starlight tends to be completely extinguished
and it is not possible to measure any polarization 
\citep[e.g.][]{1976AJ.....81..958V,1979AJ.....84..199W}

The method of tracing the magnetic field using
submillimetre polarization by preferential emission can 
typically only be measured in regions of high extinction --
A$_V$ $\sim$ 10-100 \citep[e.g.,][]{2000ApJ...537L.135W,2001ApJ...562..400M,
2006MNRAS.369.1445K}. Hence the submillimetre data can only trace
the field orientation towards the high density central regions of the clouds.
This is because these data use preferential emission from aligned dust
grains to produce the polarisation, and are only sensitive to very
high column densities in cloud centres.

At lower column densities, submillimetre 
polarimeters do not currently have the
sensitivity to detect polarized emission.
Furthermore, the nature of the usual data-taking method,
using a chopping secondary mirror, ensures that any spatially extended 
components associated with the outer parts of the cloud are chopped away.
Consequently, a combination of these two methods, of 
optical and submillimetre polarimetry, in theory provides a 
way of tracing the magnetic field from the low extinction outer parts of
molecular clouds, to the high extinction, compact, 
central, dense parts of clouds.

It has previously been found \citep{1995ApJ...448..748G} 
that near-infrared polarisation mapping 
of background stars does not accurately trace the magnetic field in dense
cloud regions. This is based on the
observation that near-infrared polarisation 
levels, caused by preferential extinction by aligned grains, do not
appear to rise sufficiently in denser cloud regions 
\citep{1995ApJ...448..748G}.

This is not an effect of extinction preventing the infrared
emission from escaping the cloud. Instead, these authors concluded that
the grains responsible for polarising background starlight are
under-represented in dense cloud regions \citep{1995ApJ...448..748G}.
There are two possible explanations for this: either there is a lack of
aligned grains in dense regions; or that only a small fraction of the dust
grain size population in dense regions lies in the correct size range to
polarise near-infrared emission from background stars.

In this paper we attempt to test 
the first of these hypotheses,
whether or not there are significant numbers of
aligned grains in high density regions.
We do this by comparing submillimetre polarization 
measurements in high-density regions with
optical polarization measurements in lower-density regions of
the same cloud for a case in which both sets of measurements
have been taken.

The submillimetre data cannot be affected by lower density foreground, or
extended, field components, since they are not sensitive to these components.
If the submillimetre and optical polarisations agree, then 
it would be difficult to avoid the conclusion that
the magnetic field is continuous from low to
high density and that the dust grains
in the high-density regions are tracing the magnetic field in those
regions. This would lead to the conclusion that
there are significant numbers of aligned grains
in the high-density regions.

\begin{figure*}
\includegraphics[angle=0,width=150mm]{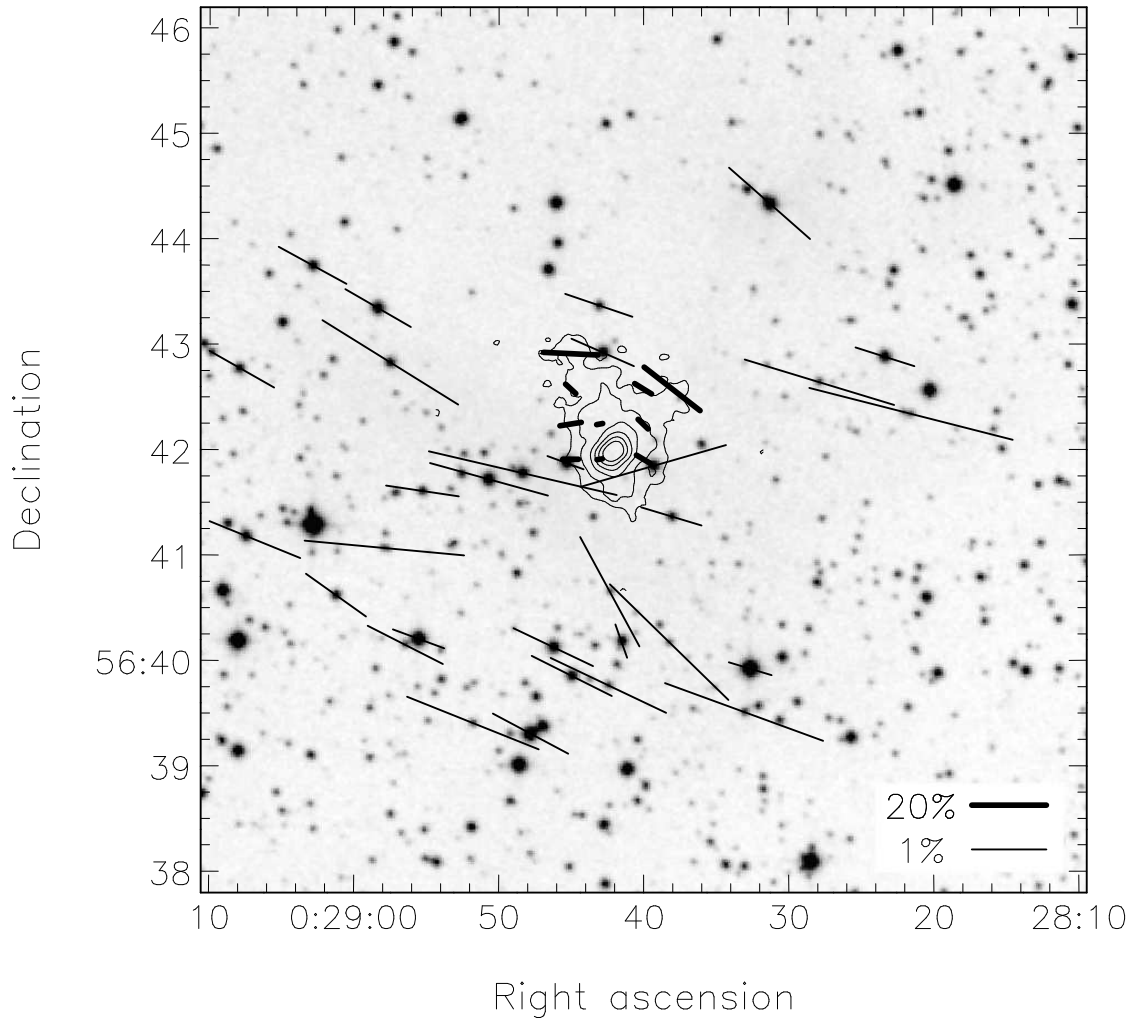}
\caption{Polarization map of CB3. The thin vectors are the optical 
polarisation vectors from \citet{2000A&AS..141..175S}. The thick 
vectors are the
submm polarisation vectors from SCUBAPOL \citep{Matthews09}, which have
been rotated through 90~degrees to illustrate the magnetic field orientation
in the same manner as the optical polarisation vectors (see text for 
discussion). The underlying greyscale image is the optical image of the 
region, taken from the digitized
sky survey (Lasker 1994), obtained using the SkyView interface
(McGlynn \& Scollick 1994).
The contours are isophotal contours of 850-$\mu$m continuum emission
as mapped by SCUBA. Note the good agreement between the field orientations
inferred independently from the optical and submm polarisation data.}
\label{cb3fig}
\end{figure*}

\section{Observations}

Optical polarization measurements were made at the 1.2-metre telescope at 
Gurushikhar, Mount Abu, India, at latitude $=$ $+$24$^o$ 36$^\prime$, 
longitude $=$ $+$72$^o$ 43$^\prime$ \citep{2000A&AS..141..175S}. The 
Cassegrain focal 
station was used, giving an $f/13$ focus. The instrument used was the
Imaging Polarimeter, IMPOL, of the Inter University Centre for Astronomy 
and Astrophysics (IUCAA), Pune, India. A description of the instrument is 
given by \citet{1994SPIE.2198..264S} and \citet{1998A&AS..128..369R}. 

The design of the IMPOL polarimeter was based on the Durham Polarimeter 
\citep{1983MNRAS.204.1163S,1985MNRAS.215..537W}. It uses a Wollaston 
prism and half-wave plate to observe two orthogonal polarization states 
simultaneously. This removes the effects of sky polarization. A CCD camera 
was used to image the data. A broad-band white light filter was used, which
corresponds to the wavelength range 0.45--0.7~$\mu$m. The observations  
were carried out with typically a 500-sec exposure time for each object.
The polarimetric data reduction procedure is described in detail by 
\citet{2000A&AS..141..175S}.  

We here report on optical polarization data of Bok globules CB3
and CB246, from the catalogue of \citet{1988ApJS...68..257C}. These data 
were first published by \citet{2000A&AS..141..175S} as part of a larger study. 
The data were taken on 1997 December 23--25. The polarization was measured of
stars background to the clouds and seen through the cloud periphery. The 
stars are generally of visual magnitude 12 or fainter. 

\begin{table}
\begin{center}
\begin{tabular}{cccccccc} \hline
No. & \multicolumn{3}{c}{R.A.} & \multicolumn{3}{c}{Dec.} &
PA \\               & \multicolumn{3}{c}{(2000)} &
\multicolumn{3}{c}{(2000)} & [$^\circ$] \\ \hline
\multicolumn{8}{c}{Submm  } \\ \hline 
-- & 00$^{\rm h}$ & 28$^{\rm m}$ & 45.1$^{\rm s}$ & 
+56$^{\circ}$ & 41$^{\prime}$ & 55$^{\prime\prime}$  & 90$\pm$19  \\
-- & 00$^{\rm h}$ & 28$^{\rm m}$ & 42.7$^{\rm s}$ & 
+56$^{\circ}$ & 41$^{\prime}$ & 55$^{\prime\prime}$  & 98$\pm$14   \\ 
-- & 00$^{\rm h}$ & 28$^{\rm m}$ & 40.3$^{\rm s}$ & 
+56$^{\circ}$ & 41$^{\prime}$ & 55$^{\prime\prime}$  & 59$\pm$8  \\ 
-- & 00$^{\rm h}$ & 28$^{\rm m}$ & 45.1$^{\rm s}$ & 
+56$^{\circ}$ & 42$^{\prime}$ & 15$^{\prime\prime}$  & 99$\pm$12 \\
-- & 00$^{\rm h}$ & 28$^{\rm m}$ & 42.7$^{\rm s}$ & 
+56$^{\circ}$ & 42$^{\prime}$ & 15$^{\prime\prime}$  & 102$\pm$14   \\ 
-- & 00$^{\rm h}$ & 28$^{\rm m}$ & 40.3$^{\rm s}$ & 
+56$^{\circ}$ & 42$^{\prime}$ & 15$^{\prime\prime}$  & 46$\pm$18   \\
-- & 00$^{\rm h}$ & 28$^{\rm m}$ & 45.1$^{\rm s}$ & 
+56$^{\circ}$ & 42$^{\prime}$ & 35$^{\prime\prime}$  & 47$\pm$23  \\
-- & 00$^{\rm h}$ & 28$^{\rm m}$ & 40.3$^{\rm s}$ & 
+56$^{\circ}$ & 42$^{\prime}$ & 35$^{\prime\prime}$  & 58$\pm$15  \\
-- & 00$^{\rm h}$ & 28$^{\rm m}$ & 37.8$^{\rm s}$ & 
+56$^{\circ}$ & 42$^{\prime}$ & 35$^{\prime\prime}$  & 52$\pm$7   \\
-- & 00$^{\rm h}$ & 28$^{\rm m}$ & 45.1$^{\rm s}$ & 
+56$^{\circ}$ & 42$^{\prime}$ & 55$^{\prime\prime}$  & 87$\pm$9   \\ \hline
\multicolumn{8}{c}{Optical  } \\  \hline
16 & 00$^{\rm h}$ & 28$^{\rm m}$ & 39.0$^{\rm s}$ & 
+56$^{\circ}$ & 41$^{\prime}$ & 19$^{\prime\prime}$ & 73$\pm$6 \\
19 & 00$^{\rm h}$ & 28$^{\rm m}$ & 49.0$^{\rm s}$ & 
+56$^{\circ}$ & 41$^{\prime}$ & 45$^{\prime\prime}$ & 76$\pm$1 \\
20 & 00$^{\rm h}$ & 28$^{\rm m}$ & 40.2$^{\rm s}$ & 
+56$^{\circ}$ & 41$^{\prime}$ & 48$^{\prime\prime}$ & 106$\pm$1 \\
21 & 00$^{\rm h}$ & 28$^{\rm m}$ & 46.1$^{\rm s}$ & 
+56$^{\circ}$ & 41$^{\prime}$ & 50$^{\prime\prime}$ & 70$\pm$2  \\
27 & 00$^{\rm h}$ & 28$^{\rm m}$ & 43.5$^{\rm s}$ & 
+56$^{\circ}$ & 42$^{\prime}$ & 52$^{\prime\prime}$ & 66$\pm$1  \\
29 & 00$^{\rm h}$ & 28$^{\rm m}$ & 43.7$^{\rm s}$ & 
+56$^{\circ}$ & 43$^{\prime}$ & 18$^{\prime\prime}$ & 71$\pm$6  \\ \hline
\end{tabular}
\end{center}
\caption{Position angles of the magnetic field vectors shown in
Figure~\ref{cb3fig}
across CB3. The submm polarisation angles
have been rotated by 90 degrees to indicate the orientation of the
magnetic field, for direct comparison with the optical polarisation
angles (see text for discussion).}
\label{cb3tab}
\end{table}

Maps of the
optical polarization vectors for these globules were mostly produced for 
the outer parts of the clouds, where the optical extinction was not so high.
CB3 exhibited polarization values 
generally below 2\% , whereas for CB246 it was a little higher. A detailed 
analysis of the nature of these polarization values was presented by 
\citet{2005MNRAS.361..177S}. 

The submillimetre observations were made at the James Clerk Maxwell
Telescope (JCMT), Mauna Kea, Hawaii, using the Submillimetre Common User
Bolometer Array (SCUBA). SCUBA contained a pair of hexagonal, close-packed,
bolometer arrays that could take data at 850 and 450 microns simultaneously,
using a dichroic beam-splitter \citep{1999MNRAS.303..659H}. SCUBA polarisation
measurements were taken with the SCUBA polarimeter, SCUBAPOL 
\citep{2003MNRAS.340..353G}, which used a rotating half-waveplate 
and fixed analyser. 

The waveplate was
stepped through 16 positions (each offset from the last by 22.5$^\circ$) and
a Nyquist-sampled image (using a 16-point jiggle pattern) was taken at each
waveplate position (Holland et al. 1996; Greaves et al. 2003).
The observations were carried out
whilst chopping the secondary mirror 120 arcsec in azimuth at 7~Hz and
synchronously detecting the signal, thus rejecting sky emission. The
integration time per point in the jiggle cycle was 1~sec, in each of the left
and right beams of the dual-beam telescope chop. The total on-source
integration time per complete cycle was 512~sec. Only 850-micron data have 
been used in this paper.
The instrumental polarization (IP) was removed using the measured value of
Greaves et al. (2003). This was 0.92$\pm$0.05\% 
at a position angle of 163$\pm$2 degrees. This was measured on planets
and found to be stable with time. The main cause of the IP
was believed to be the telescope wind blind.
SCUBA and SCUBAPOL suspended operations in 2006, pending the
commissioning of the next generation SCUBA2 instrument. 

\citet{2008ApJS..175..277D} and \citet{Matthews09} have released
near-complete legacy datasets for SCUBA and SCUBAPOL respectively. The
SCUBA polarisation data presented in this paper come from the 
\citet{Matthews09} legacy data set. 
The data were originally taken on 1998 August 10.
The Stokes I, Q \& U maps 
were calculated by \citet{Matthews09} by fitting a sinusoid
to the intensity measurements from the different waveplate angles. 
The Stokes maps were then converted by us into tabulated lists of polarisation
vectors. We binned the I, Q \& U maps to a resolution of 20 arcsec, and then
calculated the polarisation vector at each pixel. Finally, a
signal-to-noise ratio cut was applied to remove spurious vectors. The
remaining vectors are plotted in Figure 1.

The continuum submillimetre data were taken separately from the polarisation
data. For CB3 the data were downloaded from the SCUBA data
archive while the data for CB246 were taken from \citet{bokpaper}. 
The data were reduced in the usual manner 
\citep[see, for example, ][]{2007MNRAS.374.1413N} using the SURF 
software package 
\citep{SURF}. The submillimetre zenith opacity for
atmospheric extinction removal was determined by comparison with the 1.3-mm
sky opacity \citep{2002MNRAS.336....1A}.

\begin{figure*}
\includegraphics[angle=0,width=150mm]{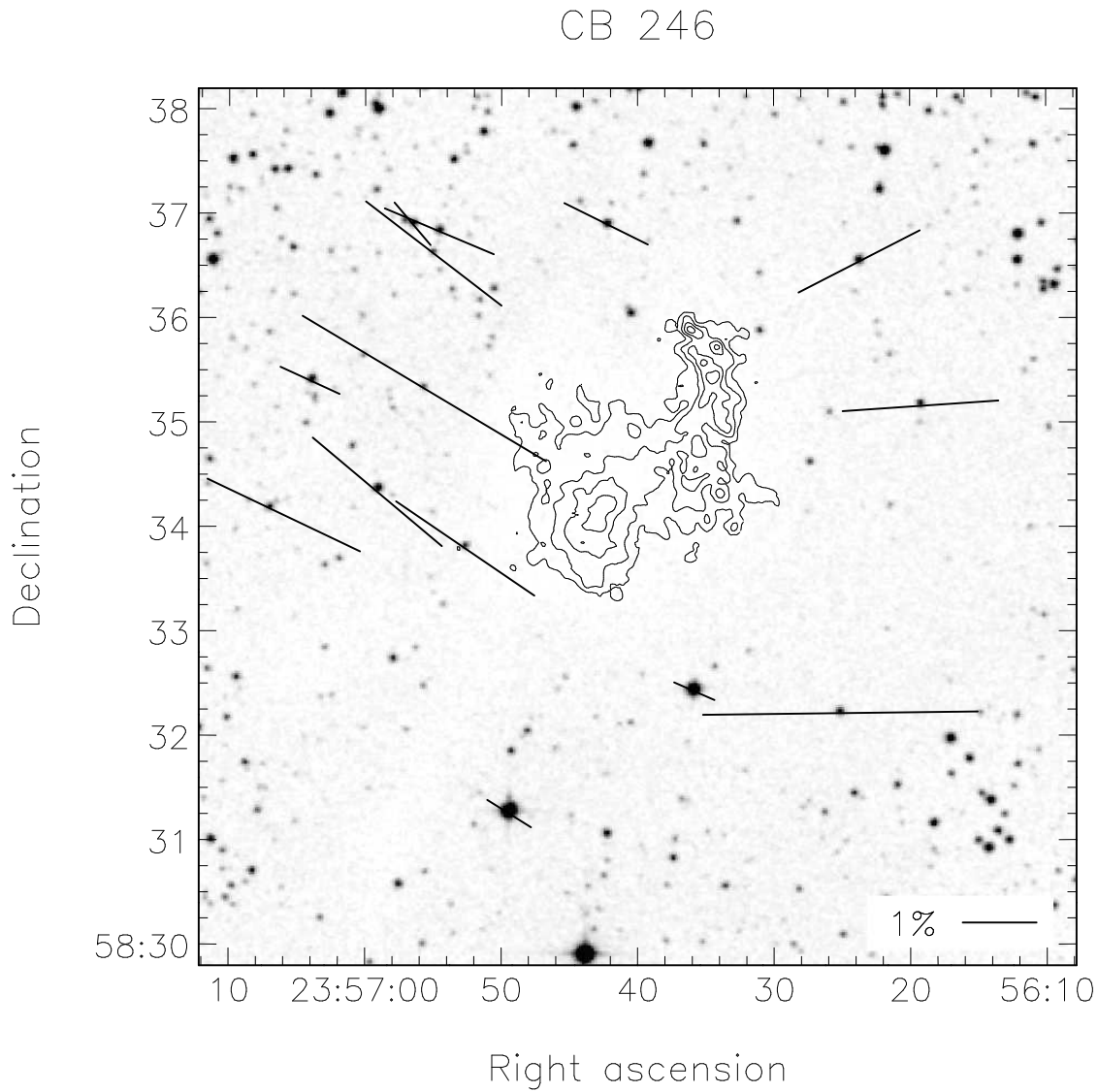}
\caption{Polarization map of CB246. Details as in Figure~1. No submm 
polarisation data were available for this region. However, note how the field
orientation inferred from the optical polarisation data lie offset from
the short axis of the submm continuum emission from the core.
Note also how both core and field appear to twist
(see text for discussion).}
\label{cb246fig}
\end{figure*}

\section{Results}

\subsection{CB3}

CB3 was catalogued by \citet{1988ApJS...68..257C},
who noted it has an angular size of 6.7$\times$5.6 arcmin. It is also 
known as Lynds 594 \citep{1962ApJS....7....1L}. It lies at an approximate 
distance of 2.5kpc
\citep{1997A&A...326..329L}. It contains at least two young stellar objects 
(YSOs), one of which is a Class 0 source \citep{1993ApJ...406..122A}
driving a chemically active bipolar outflow \citep{1999A&A...350..659C}.
Additionally there appear to be more than 20 very red K-band sources in the 
field clustered around the core \citep{1998ApJS..119...59L}. This cluster was
compared to the submillimetre emission by \citet{2000A&A...362..635H}.
CB3 is sometimes referred to as a site of intermediate
mass star formation \citep{1999A&A...350..659C,2000AGM....17..B06K}. 
It is coincident with the IRAS source 00259+5625, 
which has a luminosity of $\sim$930~L$_\odot$ \citep{1999A&A...350..659C}. 
There is also a water maser source associated with this globule 
\citep{1991MNRAS.249..763S}.

Figure~\ref{cb3fig} shows the optical and submillimetre
polarisation maps of CB3, superposed on a contour map of submillimetre
continuum emission and a grey-scale image of optical continuum emission.
The region of brightest submillimetre emission coincides with a blank
area in the optical image, as expected for a dark dust cloud.
Nevertheless, background stars are visible through the edge of the
cloud. The optical and submillimetre polarisation vectors
mostly appear to lie in a similar orientation. 

We find the E(B$-$V) colours of the
background stars to all lie in the range between 0.1 and
0.6 (Sen et al., in prep.). Taking the canonical ratio
(Whittet et al., 2008, and references therein), we therefore estimate
that these stars have A$_V$ $\sim$ 0.3--1.8.
In the region covered by
the outermost submillimetre contour, the mean column density is
$\sim$10$^{22}$cm$^{-2}$ (Launhardt et al., 1997). This corresponds
approximately to A$_V$ $\sim$ 10 (Frerking et al., 1982). This is
precisely the level of extinction in which Goodman et al., (1995)
found near-infrared polarisation levels lower than expected.

Table~\ref{cb3tab} lists the positions and position angles of the
submillimetre vectors across CB3.
The angles have been rotated by 90 degrees to show the orientation of the
magnetic field. The mean and standard deviation of the position angle of the
magnetic field derived from the Q and U values at each of the 10 positions
is 69$\pm$14 degrees. 

Table~\ref{cb3tab} also lists
the six optical magnetic field vectors which are closest to the
submillimetre vectors -- those optical vectors within an arcmin
of the centre of the submillimetre emission. The mean and standard
deviation of the position angle of these six 
optical vectors is 77$\pm$15
degrees. Vector number 20 is seen to be significantly different from
the other five vectors. This may be due to it being a
foreground star. We note that it does have a lower than
average value of E(B-V), consistent with this explanation
(Sen et al., in prep.).
If we exclude this vector and calculate the mean of 
the remainder, we obtain a mean of 72$\pm$4 degrees.

Therefore, we see that the field orientation as estimated from the 
submillimetre polarisation is in very good agreement with that
estimated from the optical polarisation. Consequently, we
deduce that the magnetic field is continuous from the low to the high
density regions, with a mean position angle of 71$\pm$14 degrees.
Furthermore, the dust grains in the high density region
appear to be tracing the magnetic field well.

The submillimetre contours on Figure~\ref{cb3fig} show that this Bok
globule is elongated. The same elongation is seen in various molecular
line tracers of this core, such as CS (Codella \& Bachiller 1999).
We estimate that the position angle of the long axis
of the globule is 21$\pm$5 degrees. This is at an angle of 50$\pm$14 degrees
to the magnetic field orientation. Models of magnetically dominated
star formation predict that the magnetic field should lie along the short 
axis of the star-forming core \citep[e.g.][]{1994ApJ...425..142C,
1998ApJ...504..280C,1996ApJ...472..211L}. In this case we see an offset of
40$\pm$14 degrees from the short axis. This is similar to results we have
seen previously in prestellar cores \citep{2000ApJ...537L.135W,
2006MNRAS.369.1445K}. We return to this in section~\ref{discus} below.

\subsection{CB246}

CB246 was catalogued by \citet{1988ApJS...68..257C},
who noted it has an angular size of 7.8$\times$4.5 arcmin. It 
is also known as Lynds 1253 \citep{1962ApJS....7....1L}. It lies 
at an approximate distance of 140pc
\citep{1997A&A...326..329L}. It is a starless core, possibly prestellar
in nature \citep{2007prpl.conf...33W}.
\citet{1996MNRAS.282..587S} list this source as inactive, with neither a 
water maser nor molecular outflow. However,
their detection of NH$_3$ and C$_2$S made them postulate that the source was 
more advanced than the other sources in their
survey. They followed up their survey with a targeted observation 
\citep{1998MNRAS.298.1092C}, which confirmed their
earlier findings. \citet{2001ApJS..136..703L} classified CB246 as having
significant red excess, but no blue asymmetry in the CS spectra,
indicating that the core has not begun to collapse.

Figure~\ref{cb246fig} shows the optical
polarisation map of CB246, superposed on a contour map of submillimetre
continuum emission and a grey-scale image of optical continuum emission.
The region of brightest submillimetre emission coincides with a blank
area in the optical image, as expected for a dark dust cloud.
Nevertheless, background stars are visible through the edge of the
cloud. No submillimetre polarimetry
data were available for this cloud, but we can infer the
plane-of-sky magnetic field orientation
from the optical polarisation vectors.

We note that the vectors to the east of the core centre have a
systematically different mean position angle from those to the west.
Similarly the core long axis orientation also appears to shift from
east to west. Therefore, we compare the mean position angles of the
magnetic field and core long axis separately in the eastern and western
parts of the core.
In the east, the plane-of-sky magnetic field lies at a position angle
of 60$\pm$3 degrees. The long axis of the core in the east
appears to lie along a position
angle of roughly 122$\pm$5 degrees. Hence the short axis is at position angle
32$\pm$5 degrees. The approximate angular offset between the magnetic field 
and the short axis of the globule is therefore 28$\pm$8 degrees for the
eastern half of the core.

In the west, the plane-of-sky magnetic field lies at a position angle
of 94$\pm$5 degrees. The long axis of the core in the west
appears to lie along a position
angle of roughly 157$\pm$5 degrees. Hence the short axis is at position angle
67$\pm$5 degrees. The approximate angular offset between the magnetic field 
and the short axis of the globule is therefore 27$\pm$10 degrees for the
western half of the core.

The weighted mean of these two offsets is therefore 28 degrees.
The mathematical error-bar on this value is less than 10 degrees.
However, given that the core and field both curve by more than 30 degrees
we do not claim an error-bar of less than 20 degrees for this estimate,
as this is the error-bar obtained if one treats the whole core simultaneously.
A detailed submillimetre polarisation map of CB246 could potentially
confirm that the curvature of the field exactly mirrors the curvature
of the core throughout its length.
In that case a smaller error-bar could be ascribed to the offset value.
Nevertheless, the value of 28 degrees
is still consistent with prestellar cores we have
observed previously (see next section).

\begin{table}
\begin{center}
\begin{tabular}{ll} 
\hline
Core & Offset \\
\hline
L1544  & 29 $\pm$ 6  \\
L183   & 34 $\pm$ 6  \\
L43    & 44 $\pm$ 6  \\
L1498  & 19 $\pm$ 12 \\
L1517B & 25 $\pm$ 6  \\
CB3    & 40 $\pm$ 14 \\
CB246  & 28 $\pm$ 20 \\
\hline
\end{tabular}
\end{center}
\caption{Position angle offsets between the magnetic field orientation
and the short axis of the core for the five prestellar cores previously
measured, together with the two sources studied in the current work
(see text for discussion).}
\label{offsets}
\end{table}

\section{Discussion}
\label{discus}

The fact that we have observed a consistent magnetic field orientation in CB3
for both the optical and submillimetre polarisation data shows that both are
tracing the same magnetic field. This implies that dust grains in the 
high-density regions of this globule are indeed tracing the magnetic field.
Hence, there must be significant numbers of aligned grains
in the high-density regions. Furthermore, the magnetic fields 
must be continuous from low to high density.
Therefore, the previous finding \citep{1995ApJ...448..748G}, that dust 
grains in high-density regions do not trace the magnetic field
in near-infrared polarisation measurements,
cannot be explained by a lack of aligned grains.
Our findings are consistent with those of Whittet et al. (2008), who
conclude from their data
that there is no change in grain alignment efficiency
from low- to high-density regions of dust clouds.

In addition, 
these data allow us to comment on the influence of magnetic fields
on the star formation process. Magnetically-dominated models of star
formation predict that a cloud should collapse first along the magnetic field
lines, and then subsequently contract quasi-statically perpendicular to
the field orientation  \citep[e.g.][]{1994ApJ...425..142C}. 
Hence these models predict that the magnetic field
should lie along the short axis of a collapsing cloud core.

\citet{2000ApJ...537L.135W} presented the first observations of the 
magnetic field geometry in pre-stellar cores when they mapped L183, 
L1544 and L43 with the SCUBA polarimeter. Their maps showed relatively 
smooth and uniform magnetic fields over the central core regions. 
However, in each case they observed an offset between the short axis of
the core and the magnetic field orientation. \citet{2006MNRAS.369.1445K} studied
the pre-stellar cores L1498 and L1517B in a similar manner and also
observed similar offsets. An attempt to explain these offsets in terms
of tri-axial cores viewed from arbitrary angles
was proposed by Basu (2000), but no magnetic model
incorporating this additional complication has been put forward.

Table~\ref{offsets} lists the cores observed previously, together with
those from this work, and the offset observed in each case between the
magnetic field orientation and the short axis of the core. The weighted mean
offset of the five previously observed cores is 31$\pm$3 degrees. We note
that the two globules observed in the current study both
have offsets consistent with this value. This is in spite of the fact
that CB3 is not a starless core, but a protostellar core.
CB246 is a starless core. The weighted mean of the six
starless cores in Table~\ref{offsets} is $\sim$30$\pm$3 degrees.
This appears to be in contradiction with some models of magnetically
moderated star formation.

\section{Conclusions}

We have used optical and submillimetre polarimetry of
Bok globule CB3 to infer the magnetic field orientation. 
The optical data trace the field orientation in low density 
regions and the submillimetre data trace
the field orientation in high density regions.
We have tested a hypothesis that there is a lack of aligned 
dust grains in high density regions, 
causing near-infrared polarisation levels to be lower than expected, by
comparing the field orientations measured by our two independent methods.

We found that the field orientation deduced from the optical polarization data
matches up well with the orientation estimated
from the submillimetre measurements.
The inferred field lines join up from low to high extinction and
produce a consistent and coherent picture of the
whole magnetic field in CB3.
We saw that both methods accurately trace the same
magnetic field. Hence we deduce
that there are significant numbers of aligned grains
in the high-density regions of CB3. Furthermore,
these aligned dust grains appear to
trace the magnetic fields in those regions,
as mapped in submillimetre polarisation.

We have also compared polarisation data with submillimetre continuum data
in Bok globules CB3 and CB246. We have compared the field orientations with
the core morphologies and found an offset in each case between the field 
orientation and the short axis of the globule. These offsets are consistent
with those seen in previous studies of prestellar cores, despite the fact
that, of our two globules, only CB246 is a starless core. Taken together,
the six starless cores that we have now studied in this way are beginning
to show a consistent picture of an offset between magnetic field orientation
and core short axis of $\sim$30$\pm$3 degrees, 
in apparent contradiction with
some models of magnetically dominated star formation.

\section*{Acknowledgements}

JMK and DJN acknowledge the UK STFC for PDRA support through the Cardiff 
Astronomy Rolling Grant.
The JCMT is operated by the Joint Astronomy Centre, Hawaii, on behalf of the 
UK STFC, the Netherlands NWO, and the Canadian NRC. The 1.2-metre telescope 
at Mount Abu is operated by the Physical Research Laboratory of India.
This research used the facilities of the Canadian Astronomy Data
Centre, operated by the National Research Council of Canada, with the
support of the Canadian Space Agency.
The Digitized Sky Survey was produced at the Space Telescope Science
Institute under U.S. Government grant NAG W-2166. The images of these
surveys are based on photographic data obtained using the Oschin
Schmidt Telescope on Palomar Mountain and the UK Schmidt Telescope at
Siding Spring. The plates were processed into the present compressed
digital form at the Royal Observatory Edinburgh (ROE) photolabs, with
the permission of these institutions.
The authors also acknowledge the use of NASA's SkyView facility
(http://skyview.gsfc.nasa.gov).

\end{document}